\documentclass[
twocolumn,
superscriptaddress,
showpacs,
english,
floatfix
]
{revtex4-2}

\usepackage[utf8]{inputenc}
\usepackage{physics}
\usepackage{graphicx}
\usepackage[dvipsnames]{xcolor}
\usepackage{microtype}
\usepackage{bm}
\usepackage{url}
\usepackage{blindtext}
\usepackage{nicematrix}

\usepackage[linktocpage=true,
  colorlinks=true, 
  pdfborder={0 0 0},
  linkcolor=blue,
  citecolor=red,
  filecolor=yellow,
  urlcolor=blue,
  bookmarks,
  pdfauthor={},
]{hyperref}
\usepackage[capitalise]{cleveref}

\newcommand{\UniBasel}{Department\ of\ Physics,\ University\ of\ Basel,\ Klingelbergstrasse\ 82,\ CH-4056\ Basel,\ Switzerland}

\newcommand{\eqeq}{E_{\text{Qeq}}}
\newcommand{\elec}{E_\text{elec}}
\newcommand{\nt}{{N_{\text{at}}}}
\newcommand{\half}{\frac{1}{2}}
\newcommand{\R}{\vb r}
\newcommand{\Q}{\vb Q}

\newcommand{\qtot}{Q_{\text{tot}}}
\newcommand{\trho}{\Tilde{\rho}(\vb G)}

\newcommand{\strain}{\varepsilon_{\mu \nu}}

\begin{document}

\title{Accelerating fourth-generation machine learning potentials\\ by quasi-linear scaling particle mesh charge equilibration}

\author{Moritz Gubler} \email{moritz.gubler@unibas.ch}          \affiliation{\UniBasel}
\author{Jonas A.\ Finkler}       \affiliation{\UniBasel}
\author{Moritz R. Schäfer} 
\affiliation{Lehrstuhl f\"ur Theoretische Chemie II, Ruhr-Universit\"at Bochum, 44780 Bochum, Germany} 
\affiliation{Research Center Chemical Sciences and Sustainability, Research Alliance Ruhr, 44780 Bochum, Germany}
\author{Jörg Behler} 
\affiliation{Lehrstuhl f\"ur Theoretische Chemie II, Ruhr-Universit\"at Bochum, 44780 Bochum, Germany} 
\affiliation{Research Center Chemical Sciences and Sustainability, Research Alliance Ruhr, 44780 Bochum, Germany}
\author{Stefan Goedecker}       \affiliation{\UniBasel}

\begin{abstract}

Machine learning potentials (MLP) have revolutionized the field of atomistic simulations by describing the atomic interactions with the accuracy of electronic structure methods at a small fraction of the costs. Most current MLPs construct the energy of a system as a sum of atomic energies, which depend on information about the atomic environments provided in form of predefined or learnable feature vectors. 
If, in addition, non-local phenomena like long-range charge transfer are important, fourth-generation MLPs need to be used, which include a charge equilibration (Qeq) step to take the global structure of the system into account. 
This Qeq can significantly increase the computational cost and thus can become the computational bottleneck for large systems.
In this paper we present a highly efficient formulation of Qeq that does not require the explicit computation of the Coulomb matrix elements resulting in a quasi-linearly scaling method. 
Moreover, our approach also allows for the efficient calculation of energy derivatives, which explicitly consider the global structure-dependence of the atomic charges as obtained from Qeq. 
Due to its generality, the method is not restricted to MLPs but can also be applied within a variety of other force fields.

\end{abstract}

\maketitle

\section{Introduction}\label{sec:intro}

Obtaining accurate potential energy surfaces (PES) at a reasonable computational expense is one of the greatest challenges in computational chemistry, physics and materials science. Since even most efficient electronic structure methods like density functional theory (DFT) are usually too demanding for large-scale simulations, the conventional approach taken in computer simulations of complex systems involves the use of heuristically derived force fields and empirical potentials. These are often able to capture the main features of the atomic interactions but are quantitatively less accurate than first principles methods based on the direct solution of the quantum mechanical equations. 

In recent years, the rapid development of data-driven machine learning potentials (MLP), which offer accurate PESs with only a fraction of the computational costs of the underlying electronic structure calculations used in the training process, has paved the way to solve this dilemma~\cite{P4885,P5673,P5793,P6102,P6121,P6131,P6112,P5366}.
A key step in the development of MLPs for large condensed systems has been the construction of the total energy as a sum of atomic energies, which only depend on the atomic environments up to a cutoff radius~\cite{2g_bp}. Many flavors of such local, second-generation MLPs~\cite{P6018} have been proposed and successfully applied to a variety of systems to date~\cite{2g_bp,P2630,P4862,P4945,P5596,P5794,P4644}. By construction they show a favorable, essentially linear scaling with system size. In addition, third-generation MLPs include long-range electrostatic interactions based on environment-dependent charges represented by machine learning~\cite{P2041,P2962,P3132,P5313,P5577}. In spite of including these long-range electrostatic interactions without truncation, such third-generation MLPs are still ``local'' in the sense that they are unable to take non-local phenomena like long-range charge transfer beyond the local atomic environments into account~\cite{P5977}.

The necessity to consider long-range charge transfer, which is present in a variety of systems, in atomistic potentials has attracted a lot of attention since several years. 
For instance, charge equilibration, which was initially developed by~\citet{qeq_91} and later refined by~\citet{NAKANO199759}, is a well-established method to approximate complicated electrostatics and charge transfer effects.
As shown by Grimme et al.~\cite{grimme_19}, charge equilibration techniques can also be used in the context of  dispersion corrections.
Warren et al.~\cite{warren_polarizability} demonstrated that the polarizability of long molecules is severely overestimated in the original Qeq approach~\cite{qeq_91} and proposed the addition of charge constraints for subsystems to overcome this problem. An overview and comparison of popular charge equilibration methods has been provided by~\citet{qeq_overview_mof}, and nowadays modern variants of Qeq are routinely used in advanced force fields such as ReaxFF~\cite{reaxff} or COMB~\cite{PhysRevB.75.085311} and are available in widely distributed simulation software packages like LAMMPS~\cite{lammps}.

The first use of charge equilibration in the framework of MLPs has been the charge equilibration neural network technique (CENT) introduced in 2015 by Ghasemi et al.~\cite{init_cent}, and this first fourth-generation MLP has been further improved in the following years~\cite{cent, cent2}. In CENT, the atomic electronegativities of the charge-equilibration approach are expressed as environment-dependent atomic properties learned by atomic neural networks with the goal to reproduce the correct total energy of the system. Due to the underlying total energy expression the CENT approach is best suited for a description of systems with primarily ionic bonding. A fourth-generation high-dimensional neural network potential (4G-HDNNP) that is more generally applicable to all types of systems has been proposed by Ko et al.~\cite{4g} by combining the advantages of CENT and second-generation HDNNPs. Here, the neural networks providing the atomic electronegativities are trained to reproduce reference atomic partial charges, and the resulting electrostatic energy is combined with modified atomic neural networks contributing atomic energies representing local bonding that explicit takes charge transfer into account.

Incorporating long-range charge transfer and the resulting electrostatics into machine learning models has become increasingly popular over the last years and many other approaches have been proposed~\cite{zubatiuk_21,Zubatyuk_21,xie_20,jacobson_22,P6666,P6612,P5933}. 
Still, so far, the application of fourth-generation MLPs has been restricted to small and medium-sized systems, primarily due to high computational cost of solving a set of linear equations which is needed in the original charge equilibration method. 
In order to improve the scaling of charge equilibration,~\citet{NAKANO199759} proposed a multilevel conjugate gradient approach that solves the set of equations iteratively. Because this approach requires the calculation of the Coulomb matrix, which is dense, the scaling is at least quadratically with respect to the number of atoms.
The performance of iterative charge equilibration schemes can be enhanced  by algorithms that do not require explicit knowledge of the matrix elements, as shown by~\citet{iter_max}. The efficiency of such 
an iterative scheme can be further improved if it is combined with the conjugate gradient method that allows to reduce the number of iterations compared to a steepest descent approach. 

Apart from the determination of the atomic charges, another important aspect that has to be considered in the development of more efficient methods is that fourth-generation MLPs often make use of the charges obtained from the charge equilibration step as input for calculating energy terms beyond simple electrostatics, which consequently also need to be taken into account in the determination of atomic forces and the stress tensor~\cite{4g}. This requires additional steps that also need to be implemented in a computationally efficient way.

In this work, we propose a formulation of the charge equilibration method, which 
combines a rapidly converging conjugate gradient with matrix times vector 
multiplications, that does not require explicit knowledge of the Coulomb matrix elements. This results in quasi-linear scaling with respect to the number of atoms in the system. Moreover, with our ansatz it is  possible to calculate derivatives of energy terms that use charges obtained from charge equilibration as input in quasi-linear time. Our method is therefore ideally suited for a combination with 4G-HDNNPs\cite{4g} and numerous other types of MLPs that depend on the atomic charges.

After a concise summary of the theoretical background in Section~\ref{sec:background}, we derive the equations for the efficient calculation of the electrostatic energy, forces and stress for periodic systems using our particle mesh charge equilibration method
in \cref{sec:method}. This approach is also expected to be very useful in other contexts requiring the solution of Poisson's equation, in case the charge density has the same form of a smooth superposition of generic atom-centered spherically symmetric charge densities. After a discussion of our results for a reference implementation in Section~\ref{sec:discussion} we conclude our main findings in Section~\ref{sec:conclusion}.

\section{Theoretical background}\label{sec:background}
\subsection{Charge Equilibration} \label{sec:qeq}

In the charge equilibration formalism the energy of the system $\eqeq$ is defined as the sum of the electrostatic energy $E_{\mathrm{elec}}$ and a Taylor expansion of atomic energies that depend on atomic charges $\mathbf{Q}=\{q_i\}$,
\begin{equation} \label{eq:eqeq}
    \eqeq(\vb R, \vb Q) = \elec(\vb R, \vb Q) + \sum_{i = 1}^{\nt} \left( E_i + \chi_i q_i +  \half J_i q_i^2 \right),
\end{equation}
with $\mathbf{R}=\{\mathbf{r}_i\}$ being the atomic coordinates and the $\{E_i\}$ the element dependent atomic reference energy offsets. These energy offsets $E_i$ cause a shift in $\eqeq$ but do not change the charges that are obtained using the charge equilibration method.
In \cref{eq:eqeq} the Taylor expansion is truncated after the second order terms and the expansion coefficients $\{ \chi_i\}$ and $\{ J_i\}$ are called electronegativity and hardness, respectively.
Moreover, the electrostatic energy is computed as $\elec = \half \int \rho(\vb r) V(\vb r) d\vb r$ and the charge density 
\begin{equation}
    \rho(\vb r) = \sum_{i = 1}^{\nt} q_i \rho_i(\norm{\vb r - \vb r_i}) \label{eq:charge_density}
\end{equation}
is a superposition of atomic charge densities $q_i \rho_i$ that are spherically symmetric around the position of atom $i$. The charge distributions $\rho_i$ are normalized to one, i.e., $\int \rho_i d\vb r = 1$, and scaled by the respective atomic charge $q_i$. $V(\vb r)$ is the electrostatic potential of the charge density $\rho$. The atomic charge densities are usually chosen to be either point charges or Gaussian charge distributions with an element specific width $\sigma_i$. 

Charge equilibration is defined as the minimization of $\eqeq$ with respect to the atomic charges $\Q$ under the constraint of keeping the total charge $\qtot$ of the system constant. Using the method of Lagrange multipliers, the  function 
\begin{equation} \label{eq:constraint}
    L = \elec + \sum_{i = 1}^{\nt} \left(E_i + \chi_i q_i +  \half J_i q_i^2 \right) + \lambda \left( \sum_{i=1}^{\nt} q_i - \qtot \right).
\end{equation}
has to be stationary.
Differentiating \cref{eq:constraint} with respect to $q_i$ and $\lambda$ yields the set of equations
\begin{align}\label{eq:dedq}
    \pdv{L}{q_i} = \pdv{\elec}{q_i} + \chi_i + J_i q_i + \lambda &= 0 \\
    \pdv{\eqeq}{\lambda} = \sum_{i=1}^{\nt} q_i - \qtot &=0. \label{eq:dedlambda}
\end{align}
Inserting the definition of the charge density into the electrostatic energy yields
\begin{equation}\label{eq:elec}
    \elec = \half \sum_{i,j = 1}^{\nt} q_i q_j \underbrace{\int \int \frac{\rho_i(\norm{\vb r - \vb r_i}) \rho_j(\norm{\vb r' - \vb r_j})}{\norm{\vb r - \vb r'}} \mathrm{d} \vb r \mathrm{d}\vb r'}_{=A_{ij}}
\end{equation}
The matrix $\vb A$ is symmetric and positive definite because $\elec = \half \vb Q^T \vb A \vb Q \geq 0$ and $\elec = 0 \iff \vb Q = 0$. The first inequality holds because the electrostatic interaction energy of a smooth charge density is always positive.
Because of the spherical symmetry of the Gaussian density $\rho_i$ around atom $i$, $A_{ij}$ is a function of the distance between $\vb r_i$ and $\vb r_j$.
Using the definition from \cref{eq:elec,eq:dedq} and $\pdv{\elec}{q_i} = \sum_{j=1}^\nt A_{ij} q_j$, the derivative of the energy with respect to the atomic charges can be simplified to
\begin{equation}\label{eq:dedq_reduced}
    \pdv{\eqeq}{q_i} = \sum_{j=1}^{\nt} A_{ij} q_j + \chi_i + J_i q_i + \lambda = 0 \quad.
\end{equation}
\cref{eq:dedq,eq:dedlambda} can be written in matrix notation
\begin{equation}\label{eq:qeq_matrix}
    \begin{pNiceArray}{ccc|c}
         & & & 1 \\
         & \vb M& & \vdots \\
         & & & 1 \\
        \hline
         1 &\dots& 1 & 0
    \end{pNiceArray}
\cdot
\begin{pNiceArray}{c}
    q_1 \\
    \vdots \\
    q_{\nt} \\
    \hline
    \lambda
\end{pNiceArray}
= 
\begin{pNiceArray}{c}
    -\chi_1 \\
    \vdots \\
    -\chi_\nt \\
    \hline
    \qtot
\end{pNiceArray}
\end{equation}
where
\begin{equation}
    M_{ij} = 
    \begin{cases}
        A_{ij}\quad \text{if} \quad i\neq j \\
        A_{ii} + J_i \quad \text{if} \quad i = j.
    \end{cases}    
\end{equation}
 This set of linear equations can be solved either directly with cubic scaling or using the iterative multilevel conjugate gradient approach\cite{NAKANO199759,iter_max}. Since the matrix $\vb{A}$ has to be calculated, the best scaling using the latter approach is $\mathcal{O}(\nt^2)$ because the Coulomb matrix is dense and has $\nt^2$ elements.

 \subsection{Calculation of total derivatives}
 
Since the charge equilibration energy depends on the atomic positions, both explicitly and implicitly, via the dependence of the charges on the atomic positions, the gradient is given by the expression 
\begin{equation}
\dv{E_{\mathrm{elec}}(\vb R, \vb Q(\vb R))}{\vb R} = \pdv{E_{\mathrm{elec}}(\vb R, \vb Q)}{\vb R} + \pdv{E_{\mathrm{elec}}}{\vb Q} \pdv{\vb Q}{\vb R}.
\end{equation}
Since in charge equilibration the energy $\eqeq$ is minimized with respect to the $q_i$, we have $\pdv{\eqeq}{\vb Q} = 0$, 
and hence the total derivative simplifies to
\begin{equation}
\dv{\eqeq(\vb R, \vb Q(\vb R))}{\vb R} = \pdv{\eqeq(\vb R, \vb Q)}{\vb R} .
\end{equation}
The situation is more complicated if the charges are not determined by a minimization of the total energy. For example in 4G-HDNNPs~\cite{4g} the charges obtained by the charge equilibration are used as parameters to calculate the short range and electrostatic energies. In that case, the gradient of the total energy of the system is given by
\begin{equation}
\dv{E(\vb R, \vb Q(\vb R))}{\vb r_i} = \pdv{E}{\vb r_i} + \sum_{j=1}^{\nt} \pdv{E}{q_j} \pdv{q_j}{\vb r_i}
\end{equation}
and the derivatives $\pdv{q_j}{\vb r_i}$ are required. In total, there are $3\nt ^2 $ derivatives of atomic charges with respect to atomic positions, which would result in an at least quadratic scaling when the derivatives are calculated. However, the total derivative $\dv{E}{\mathbf{r}_i}$ can be evaluated without calculating $\pdv{q_j}{\mathbf{r}_i}$ explicitly. A similar approach has been used by ~\citet{deriv_trick_19} in the context of polarizable force fields.~\citet{4g} derived an efficient way to calculate the forces in the 4G-HDNNP method. The resulting formulas for the forces and the strain derivatives $\dv{}{\strain}$ needed for calculating the stress are 
\begin{align}
    \dv{E}{\vb r_k} &= \pdv{E}{\vb r_k} + \sum_{i=1}^\nt \lambda_i \left( \sum_{j=1}^\nt \pdv{A_{ij}}{\vb r_k}  q_j + \pdv{\chi_i}{\vb r_k}\right) \label{eq:force_trick} \\
    \dv{E}{\strain} &= \pdv{E}{\strain} + \sum_{i=1}^\nt \lambda_i \left( \sum_{j=1}^\nt \pdv{A_{ij}}{\strain}  q_j + \pdv{\chi_i}{\strain}\right) \label{eq:straintrick}
\end{align}
where $\bm \lambda$ can be obtained by solving
\begin{equation} \label{eq:lambda_equilibration}
    \vb M \bm \lambda = - \pdv{E}{\vb Q}
\end{equation}
under the constraint that the sum of the components of $\bm \lambda$ is 0.

In the derivation of the total derivatives it is assumed that the electronegativities $\chi_i$ depend on the atomic positions whereas  the hardnesses $J_i$  are only element-specific quantities. While the derivation could also be extended to consider position-dependent hardnesses, we assume that the hardness is an element-specific parameter throughout this paper.

The cost of evaluating \cref{eq:force_trick,eq:straintrick} directly still scales at least quadratically. In \cref{sec:qeq_derivative_iter} it will be discussed how these sums can be evaluated more efficiently.

\section{Particle Mesh Charge Equilibration}\label{sec:method}

\subsection{Solving the system of equations}

We start by noting that, as shown in \cref{asec:dedq}, $\pdv{\elec}{q_i}$ can be written as
\begin{equation} \label{eq:mesh_dedq}
    \pdv{\elec}{q_i} = \int \rho_i(\norm{\vb r - \vb r_i}) V(\vb r) d\vb r = \left(\vb A \cdot \vb Q\right)_i.
\end{equation}
With this relation, the matrix-vector product $\vb A \cdot \vb Q$ can be calculated for an arbitrary $\vb Q$ without any explicit knowledge about the elements of matrix $\vb A$. Being able to calculate matrix-vector products for arbitrary vectors is sufficient to solve a system of equations iteratively.

As discussed in \cref{sec:qeq}, the Coulomb matrix is positive definite. The manifold of the charge equilibration constraint $\sum_i q_i = Q_{\mathrm{tot}}$ is a hyperplane meaning that an arbitrary large step along a constrained gradient will still fulfil the charge conservation constraint. Therefore, the standard conjugate gradient method can be used to solve the set of linear equations 
\begin{eqnarray}
(\vb A\cdot \vb Q)_i + J_i q_i = (\vb M \cdot \vb Q)_i = - \chi_i
\end{eqnarray}
where 
\begin{eqnarray}
(\vb M \cdot \vb Q)_i = \pdv{E}{q_i} + J_i q_i.
\end{eqnarray}

The constrained gradient $\widehat{\pdv{\elec}{q_i}}$ of \cref{eq:mesh_dedq} can be obtained by projecting the gradient onto the constraint
\begin{eqnarray}
\widehat{\pdv{\elec}{q_i}} = \pdv{\elec}{q_i} -\frac{1}{N_{\mathrm{at}}} \sum_{j=1}^{N_{\mathrm{at}}} \pdv{\elec}{q_j}.
\end{eqnarray}

\subsection{Plane wave methods}

In order to minimize $\eqeq$ fast, it is necessary to evaluate \cref{eq:mesh_dedq} efficiently. In the case of periodic boundary conditions this can be done by solving Poisson's equation in Fourier space using plane waves.
Let $\Tilde{\rho}(\vb G)$ and $\Tilde{V}(\vb G)$ be the Fourier transforms of $\rho$ and $V$, respectively and $\vb G$ is a Fourier space vector. Because of the Plancherel theorem, $\elec$ can be calculated in Fourier space as
\begin{equation}\label{eq:elec_fourier}
    \elec = \half \int \rho(\vb r) V(\vb r) d\vb r = \half \int \Tilde{\rho}^* (\vb G) \Tilde{V}(\vb G) d\vb G
\end{equation}
where the superscript $^*$ of $\rho$ represents complex conjugation. 
This is particularly useful, because Poisson's equation can be solved analytically in Fourier space with the solution $\Tilde{V}(\vb G) = -4 \pi \frac{\Tilde \rho(\vb G)}{\vb G^2}$. The electrostatic energy can be calculated by Fourier transforming $\rho$ and then solving the Fourier space integral in \cref{eq:elec_fourier}.
Then, the electrostatic potential $V(\vb r)$ in real space can be obtained efficiently by back transforming the Fourier coeficients $-4 \pi \frac{\Tilde \rho(\vb G)}{\vb G^2}$.

In case of periodic boundary conditions, the integral in \cref{eq:elec_fourier} transforms into the following series
\begin{equation}
    \elec = 2 \pi \Omega \sum_{\vb G} \Tilde{\rho}^* (\vb G) \frac{\trho}{\vb G^2}
\end{equation}
where $\Omega$ is the unit cell volume. Also, Fourier transforms of any periodic function can be obtained numerically using the Fast Fourier transform (FFT), which can be calculated with a $\mathcal{O}(N \ln N)$ scaling where $N$ the number of gridpoints.

The electrostatic potential $V(\vb r)$ can be obtained using a forward and backward Fourier transform. The atomic charge densities $\rho_i$ which are present in \cref{eq:mesh_dedq} typically decay exponentially, which makes it possible to obtain $\pdv{\elec}{q_i}$ for all atoms in quasi linear time because it is sufficient to integrate only over a small volume around each atom $i$ to obtain $\pdv{\elec}{q_i}$. Therefore, the matrix vector product $(\vb M \cdot  \vb Q)_i = \pdv{E}{q_i} + J_i q_i$ can be evaluated for any $\vb Q$ in quasi linear time.

\subsection{Derivatives with plane wave methods}

Most materials science simulations require the calculation of the forces acting on the nuclei and the stress tensor acting on the periodic lattice. Using the definition of $\rho$ from \cref{eq:charge_density}, the electrostatic force can be obtained by evaluating the real space integral
\begin{equation}\label{eq:elec_force}
    \pdv{\elec(\vb r_1, \dots, \vb r_{N_{\mathrm{at}}})}{\vb r_i} = q_i \int V(\vb r) \pdv{\rho_i(\norm{\vb r - \vb r_i)}}{\vb r_i} d \vb r
\end{equation}
which is derived in \cref{asec:forces}.

Regarding the electrostatic stress, two different definitions of stress are commonly in use, the microscopic stress and the macroscopic stress.
The microscopic stress tensor is a tensor field, and an example is the Maxwell stress
$\sigma_{ij} = \frac{1}{4 \pi} \left(  E_iE_j - \half \delta_{ij}E^2 \right)$ for systems without periodic boundary conditions.
For periodic systems, the microscopic stress tensor is rarely used since in bulk materials it is often sufficient to consider the macroscopic stress tensor, which corresponds to the average microscopic stress per unit volume. The symmetric strain tensor $\strain$ describes an infinitesimal deformation of a crystal $r'_\mu = (\delta_{\mu \nu} + \strain) r_\nu$ where $\delta$ stands for the Kronecker delta and the Einstein summation convention is used.
The macroscopic stress $\bm{\sigma}$ is the strain derivative of the total energy per unit volume\cite{stress_1,stress2}  with
\begin{equation}\label{eq:strain_derivative}
\sigma_{\mu \nu} = \frac{1}{\Omega} \left. \pdv{E(\vb R ')}{\strain}\right|_{ \varepsilon = 0} .
\end{equation}
$\vb R'$ contains the atomic positions $\vb R$ that are deformed with the strain tensor $\strain$.
Because of the variational character of electronic structure calculations, the strain derivative of the charge density is zero, which reduces \cref{eq:strain_derivative} to the average Maxwell stress that can easily be calculated when the total potential $V(\vb r)$ is known. The strain derivative of the charge density is not zero in our case and the total stress has to be evaluated in Fourier space where $\pdv{\rho}{\strain}$ is given by
\begin{widetext}
\begin{equation}\label{eq:reciprocal_stress}
    \sigma_{\mu \nu} = \frac{1}{\Omega} \left. \pdv{\elec(\vb R')}{\strain}\right|_{\varepsilon=0} = \delta_{\mu \nu} \frac{E}{\Omega}
    + 2 \pi \sum_{\vb G\neq 0} \left( \frac{1}{\vb G^2}  \left(\Tilde \rho^*(\vb G) \pdv{\Tilde \rho (\vb G)}{\strain} + \Tilde \rho(\vb G) \pdv{\Tilde \rho ^* (\vb G)}{\strain} \right) 
    + 2 \frac{\abs{\Tilde{\rho}(\vb G)}^2}{\vb G^4} G_\mu G_\nu \right).
\end{equation}
A derivation of \cref{eq:reciprocal_stress} can be found in \cref{asec:stress}.
The Fourier transform of the strain derivative $\pdv{\Tilde \rho (\vb G)}{\strain}$ can be obtained by transforming the strain derivatives $\pdv{\rho}{\strain}$ of the charge density into Fourier space, which requires six additional Fourier transforms as $\pdv{\rho}{\strain}$ is symmetric.
The periodic generalization of \cref{eq:charge_density} for the lattice matrix $\vb h$ containing the three lattice vectors $\vb h_1$, $\vb h_2$ and $\vb h_3$ is given by
\begin{equation} \label{eq:rho_pbc}
    \rho(\vb r, \vb r_1, \dots, \vb r_{N_{\mathrm{at}}}, \vb h) = \sum_{i, j, k = -\infty}^\infty \sum_{l=1}^{N_{\mathrm{at}}} q_l \rho_l\left( \norm{ \vb r - \vb r_l - i \vb h_1 - j \vb h_2 - k \vb h_3  }^2 \right).
    \end{equation}
The calculation of the strain derivative of \cref{eq:rho_pbc} is discussed in \cref{asec:strain}. It has the form
\begin{align}
\label{eq:rho_pbc2}
        & \pdv{\rho(\R, \R_1, \dots, \R_{N_{\mathrm{at}}}, \vb h)}{\strain} = \nonumber\\
        & \sum_{i, j, k = -\infty}^\infty \sum_{l=1}^{{N_{\mathrm{at}}}} 2  q_l ( \vb r - \vb r_l - i \vb h_1 - j \vb h_2 - k \vb h_3 )_\mu ( \vb r - \vb r_l - i \vb h_1 - j \vb h_2 - k \vb h_3 )_\nu \rho_l'(( \vb r - \vb r_l - i \vb h_1 - j \vb h_2 - k \vb h_3 )^2).
\end{align}
The sums over $i$, $j$ and $k$ in \cref{eq:rho_pbc,eq:rho_pbc2} are over all periodic images of the simulation cell.

\subsection{Particle mesh total derivatives} \label{sec:qeq_derivative_iter}

Using the definitions
\begin{equation}
    \rho^{\bm{\lambda}} = \sum_{i=1}^\nt \lambda_i \rho_i \quad 
    \text{and} 
    \quad \rho^{\vb Q} = \sum_{i=1}^\nt q_i \rho_i,
\end{equation}
the corresponding potentials $V^{\vb{Q}}$ and $V^{\bm \lambda}$ as well as the electrostatic energies 
\begin{equation}
E^{\bm \lambda} = \half \int \rho^{\bm \lambda} V^{\bm \lambda} d\vb r \quad
    \text{and} \quad
    E^{\vb Q} = \half \int \rho^{\vb Q} V^{\vb Q} d\vb r
\end{equation}
the double sums in \cref{eq:force_trick,eq:straintrick} can be expressed in terms of the newly introduced variables as
\begin{align}
    \sum_{i, j=1}^\nt \lambda_i \pdv{A_{ij}}{r_k}  q_j &= \int \left[ V^{\vb Q}(\vb r) \pdv{\rho^{\bm \lambda}(\vb r)}{\vb r_k} + V^{\bm \lambda}(\vb r) \pdv{\rho^{\vb Q}(\vb r)}{\vb r_k} \right] d\vb r \label{eq:force_int}
\end{align}
and
\begin{align}
    \sum_{i,j=1}^\nt \lambda_i \pdv{A_{ij}}{\strain}  q_j &= \sum_{i, j=1}^\nt \int q_i \lambda_j \pdv{}{\strain}\frac{\rho_i(\vb r) \rho_j(\vb r')}{\norm{\vb r - \vb r'}} d\vb r d\vb r' \label{eq:strain_part1}
\end{align}
\cref{eq:force_int} can be solved in real space and \cref{eq:strain_part1} can be solved in Fourier space, in which the integral has the  form
\begin{align}\label{eq:reciprocal_qeqstrain}
    \sum_{i,j=1}^\nt \lambda_i \pdv{A_{ij}}{\strain}  q_j 
    =2 \pi \Omega \sum_{\vb G\neq 0} \frac{1}{\vb G^2} \left[ \delta_{\mu\nu}  \rho^{\bm{\lambda} *}(\vb G) \Tilde \rho^{\vb Q}(\vb G) +  \pdv{ \tilde \rho^{\bm \lambda *}(\vb G)}{\strain} \tilde \rho^{\vb Q}(\vb G) + \tilde \rho^{\bm \lambda*}(\vb G) \left( \pdv{ \tilde \rho^{\vb Q}(\vb G)}{\strain} + \frac{2}{\vb G^2} G_\mu G_\nu  \tilde \rho^{\vb Q}(\vb G)\right) \right]
\end{align}
\end{widetext}
which can all be evaluated with quasi-linear scaling. A derivation of \cref{eq:force_int,eq:reciprocal_qeqstrain} is presented in \cref{asec:qeq_tot_deriv,aq:4g_stress_trick} in Appendix \ref{asec:totderiv}.

\begin{figure}[t]
    \centering
    \includegraphics[width=\columnwidth]{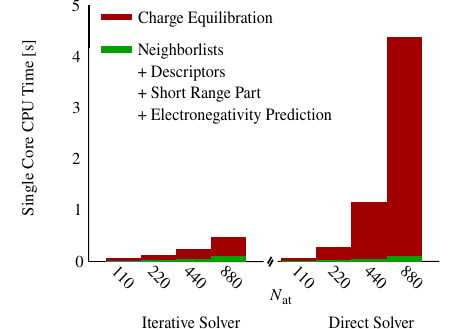}
     \caption{
       Benchmark results illustrating the performance of the iterative particle mesh electrostatic method and the conventional direct method.
       The figure displays the average single core CPU times required for predicting the energies of 50 randomly selected periodic structures from a dataset containing Au$_2$ clusters on undoped and doped MgO(001) surfaces described by Ko \textit{et al.}~\cite{4g}. 
       Each structure comprises 110 atoms, along with their respective supercells containing 220, 440, and 880 atoms.
       The benchmark includes both the direct method employed in their study on the right and the newly introduced iterative particle mesh method on the left.
       The simulations were conducted using the RuNNer code~\cite{P4444,P5128} on a single core CPU on a Intel Xeon 6430 processors with 32 cores, operating at 2.10 GHz and a 270W TDP.
       The system is equipped with 512\,GB DDR5-4800 ECC REG RAM (16$\cdot$32\,GB). 
     }
    \label{fig:4g_plot}
\end{figure}

The coefficients $\lambda_i$ can be obtained by solving $\vb M \cdot \bm \lambda = - \pdv{E}{\vb Q}$, where $E$ is an arbitrary energy that depends on charges obtained by minimizing $\eqeq$. Once again, the matrix-vector product $(\vb A \cdot \bm \lambda)_i = \pdv{E^{\bm \lambda}}{\lambda_i} = \int \rho_i(\norm{\vb r - \vb r_i}) V^{\bm \lambda} d\vb r$ can be evaluated without knowledge of the elements of $\vb A$. Therefore, it can also be solved iteratively using the conjugate gradient method.
\begin{figure}[t]
    \centering
    \includegraphics{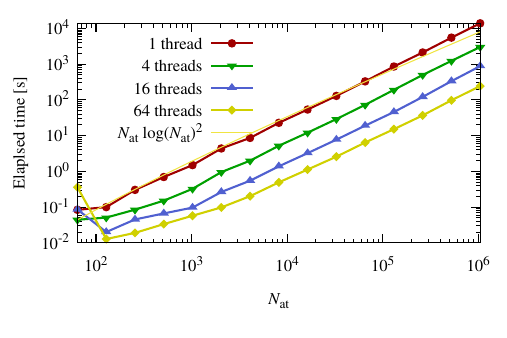}
     \caption{Timings of our reference implementation of the iterative particle mesh charge equilibration. Shown is the time calculate the Qeq charges for a periodic system as a function of the number of atoms in the system for different numbers of OPENMP threads. The calculations were performed on an AMD EPYC 7742 64-Core Processor with one TB of RAM. Less than 30\,GB of  RAM were used in total for all threads for all benchmark calculations. The fitted yellow line shows the asymptotic scaling of the computational cost of the newly developed iterative method.
     }
    \label{fig:scaling_plot}
\end{figure}

\section{Results and Discussion}\label{sec:discussion}

Overall, our iterative particle mesh charge equilibration method can be summarized as follows:

\begin{enumerate}
    \item  \textit{Charge equilibration:} Solve $\vb M \cdot \vb Q = -\bm \chi$ under the constant charge constraint using the conjugate gradient method. $(\vb M \cdot \vb Q)_i = \pdv{\elec}{q_i} + J_i q_i$.
    
    \item \textit{Calculate $\bm \lambda$ charges:} Solve $\vb M \cdot \bm \lambda = - \pdv{E(\vb r_1, \dots, \vb r_\nt \vb Q)}{\Q}$ under the constraint that $\sum_i \lambda_i = 0$ using the conjugate gradient method.
    
    \item \textit{Calculate forces and stress:} Use \cref{eq:force_trick,eq:straintrick,eq:force_int,eq:reciprocal_qeqstrain} to calculate the forces and the stress tensor.
\end{enumerate}

In a first step we have compared the computational efficiency of the particle mesh charge equilibration method with the conventional charge equilibration approach, i.e., the direct solution of a set of linear equations.
For this purpose, the presented iterative particle mesh electrostatic method has been incorporated into our MLP software RuNNer~\cite{P4444,P5128} yielding a significant enhancement for molecular simulations employing 4G-HDNNPs.
Results, comprehensively presented in Fig. \ref{fig:4g_plot}, show large performance gains of the new iterative electrostatic approach compared to the conventional direct method.
Examination of the computation time distribution reveals the significant contribution of the electrostatic component, highlighting its dominance in the overall computational costs.

Having confirmed the high performance of our new method with respect to the conventional approach, we now turn to the scaling behavior of the method.
Assuming that the number of iterations needed in the conjugate gradient method does not depend on the system size, the computational costs for the iterative charge equilibration are determined by the cost of solving the electrostatic problem, which scales like $\mathcal{O}(\nt \ln \nt)$ in our case. 
The number of conjugate gradient iterations needed to converge the charge equilibration charges is slightly increasing with the number of particles. Therefore, in reality the asymptotic behaviour to some extent deviates from the ideal $\mathcal{O}(\nt \ln \nt)$ scaling. 
Therefore, the charge equilibration and the computation of the derivatives have a quasi linear scaling which is slightly higher than $\mathcal{O}(\nt \ln \nt)$. \cref{fig:scaling_plot} shows timings of our reference implementation of the iterative charge equilibration method presented in this paper. The asymptotic scaling appears to be approximately $\mathcal{O}(\nt \ln(\nt)^2)$. 
The required calculation of the charge density and the 3-dim FFT's can be efficiently parallelized 
with OpenMP resulting in good overall parallel speed-ups.

\begin{figure}[t]
    \centering
    \includegraphics{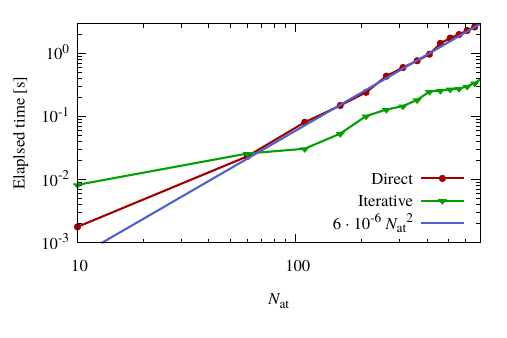}
    \caption{Logarithmic plot of the iterative particle mesh charge equilibration timings in comparison with the standard direct approach employing a solution of the set of linear equations. The Coulomb matrix elements were calculated using standard Ewald techniques for the direct approach, which is governing the scaling for the shown small system sizes resulting in dominantly quadratic scaling here. The calculation was done on a desktop machine with an 11th generation Intel i7 CPU (11700) and 32\,GB of RAM. The fitted blue line shows the scaling of the computational cost of the direct method.}
    \label{fig:log_compare}
\end{figure}

In \cref{fig:log_compare} the particle mesh iterative solver is compared with the standard direct approach to identify at which system size the new approach becomes more efficient. We find that the iterative method developed in this paper is faster than the direct solution when the test system exceeds a size of about 100 atoms. 
The computational cost of obtaining the solution for the conventional direct approach scales cubically because of the system of equations to be solved. Since there are highly efficient solvers available for  systems of linear equations, the prefactor of the cubic term is small and the quadratic term is dominant in the relatively small system size range shown in \cref{fig:log_compare}, while the cubic scaling is anticipated to become dominant for systems between 1000 and 10000 atoms. The quadratic term in the direct approach results from calculating the matrix elements of $\vb A$. Our newly proposed iterative method therefore outperforms the direct approach even for system sizes where the cost of solving the system of equations is negligible and the most expensive part in the direct method is calculating all the matrix elements $A_{ij}$.
Being iterative, our approach can profit from a good input guess to reduce the number of iterations. This effect was not taken into account in our tests but will exist in many real applications. In Molecular Dynamics (MD) for instance, the charges from the previous MD step form a good input guess for the next MD step. This will further increase the efficiency gains of our iterative method compared to the standard direct method.

\section{Conclusions} \label{sec:conclusion}

In this work we have presented a quasi-linear, i.e. $N\log (N)^2$, scaling method for charge equilibration that allows us to speed up the evaluation of any atomistic potential that contains a charge equilibration part. The 
atomistic potential can either be a machine learning potential or a classical force field. The performance of our method has been investigated for the example of a fourth-generation high-dimensional neural network potential.
We have shown that due to the high efficiency of the method it is now possible to perform simulations of systems containing thousands of atoms, which to date has been very demanding if  long-range charge transfer has to be taken into account.
Consequently, our method will allow to treat even complex systems with the latest generation of machine learning potentials to enable simulations of unprecedented accuracy.

\begin{acknowledgments}

 Financial support was obtained from the Swiss National Science Foundation (project $200021\_191994$) and the Deutsche Forschungsgemeinschaft (DFG, German Research Foundation, project 495842446) in the framework of the DFG priority program SPP 2363. This work has been supported by the DFG under Germany’s Excellence Strategy—EXC 2033–390677874—RESOLV. The calculations were performed on the computational resources of the Swiss National Supercomputer (CSCS) under project s1167 and on the Scicore (\url{http://scicore.unibas.ch/}) computing center of the University of Basel.
\end{acknowledgments}

\appendix
\begin{widetext}

\section{Derivation of charge derivatives}\label{asec:dedq}
Using the definition of $\elec$ given in \cref{eq:elec}, $\pdv{\elec}{q_i}$ can be derived the following way:
\begin{align}
    \pdv{\elec}{q_i} & = \half \int \int \sum_{j, k = 1}^\nt \frac{\rho_j (\vb r - \vb r_j) \rho_k(\vb r' - \vb r_k)}{\norm{\vb r - \vb r'}} \pdv{}{q_i} q_j q_k d\vb r d\vb r'
    = \int  \rho_i(\vb r- \vb r_i) \underbrace{\int \frac{\sum_{j=1}^{\nt} q_j \rho_j(\vb r' - \vb r_j)}{\norm{\vb r - \vb r'}} d\vb r'}_{=V(\vb r)} d\vb r \nonumber\\
    &= \int V(\vb r) \rho_i(\vb r - \vb r_i) d\vb r
\end{align}

\section{Derivation of electrostatic forces}\label{asec:forces}
\begin{equation} \label{aq:electrostatic_forces}
    \pdv{E(\vb R_1, \dots, \vb R_{N_{\mathrm{at}}})}{\vb R_i} = \half \int \pdv{}{\vb R_i} \left(\rho(\vb r) V(\vb r) \right) d\vb r = \half \int \left[ V(\vb r) \pdv{\rho(\vb r)}{\vb R_i} + \rho(\vb r) \pdv{V(\vb r)}{\vb R_i} \right] d\vb r = \int V(\vb r) \pdv{\rho(\vb r)}{\vb R_i} d \vb r.
\end{equation}
The last part of \cref{aq:electrostatic_forces} holds because 
\begin{equation} \label{aq:realspace_forces}
    \int \rho(\vb r) \pdv{}{\vb R_i} \underbrace{\int \frac{\rho(\vb r')}{\norm{\vb r - \vb r'}}d\vb r'}_{=V(\vb r)} d\vb r = \int \int \frac{\rho (\vb r) \pdv{\rho(\vb r')}{\vb R_i}}{\norm{\vb r - \vb r'}} d\vb r' d\vb r = \int \pdv{\rho (\vb r')}{\vb R_i} \underbrace{\int \frac{\rho(\vb r)}{\norm{\vb r - \vb r'}} d\vb r}_{=V(\vb r')} d\vb r' = \int V (\vb r) \pdv{\rho(\vb r)}{\vb R_i} d \vb r.
\end{equation}
$\pdv{\rho}{\vb R_i} = q_i \pdv{\rho_i}{\vb R_i}$ when the definition of $\rho$ from \cref{eq:charge_density} is used. Therefore, 
\begin{equation}
    \pdv{E(\vb r_1, \dots, \vb r_{N_{\mathrm{at}}})}{\vb r_i} = q_i \int V(\vb r) \pdv{\rho_i(\vb r -\vb r_i)}{\vb r_i} d \vb r. \label{aq:forces}
\end{equation}

\section{Calculation of electrostatic stress} \label{asec:stress}
The stress tensor of the electrostatic energy is derived for periodic systems by calculating the strain derivative of the electrostatic energy in Fourier space because it requires the least number of additional Fourier transforms.
\begin{equation}
    \pdv{E}{\strain} = \pdv{}{\strain} 2 \pi\Omega \sum_{\vb G\neq 0} \frac{\abs{\Tilde{\rho}(\vb G)}^2}{\vb G^2} = 2 \pi \underbrace{\pdv{\Omega}{\strain}}_{\vb A} \sum_{\vb G\neq 0} \frac{\abs{\Tilde{\rho}(\vb G)}^2}{\vb G^2} + 2 \pi\Omega
    \sum_{\vb G\neq 0} \left( \frac{1}{\vb G^2} \underbrace{\pdv{\abs{\Tilde{\rho}(\vb G)}^2}{\strain}}_{\vb B} + \abs{\Tilde{\rho}(\vb G)}^2 \underbrace{\pdv{}{\strain} \frac{1}{\vb G^2}}_{\vb C} \right)
\end{equation}
The following identities are useful for calculating the strain derivatives present in the terms $\vb A$, $\vb B$ and $\vb C$:
\begin{align}
    & \text{Strain derivative of a position vector } &\pdv{r_\tau}{\strain} &= \delta_{\tau \mu} r_\nu \label{aq:real_vec}\\
    & \text{Strain derivative of a Fourier space vector} & \pdv{G_\tau}{\strain} &= -\delta_{\tau \nu} G_\mu \label{aq_rec_lat} \\
    & \text{Strain derivative of $\frac{1}{\vb G^2}$} & \pdv{}{\strain} \frac{1}{\vb G^2} & = \frac{2}{\vb G^4} G_\mu G_\nu \label{eq:strain_green} \\
    & \text{Strain derivative of the unit cell volume} & \pdv{\Omega}{\strain} & = \delta_{\mu \nu} \Omega \label{eq:strain_omega} \\
    & \text{Strain derivative of a product of a real space and a Fourier space vector} & \pdv{}{\strain}r_\tau G_\tau & = 0 \label{eq:fourier_factor}
\end{align}
The Einstein summation convention is used in all of the identities. \cref{aq:real_vec} can be derived from the definition of the translation $r_\mu ' = (\delta_{\mu \alpha} + \varepsilon_{\mu\alpha} ) r_\alpha$. For \cref{aq_rec_lat}, the first Taylor expansion coefficient of the inverse transformation $((\delta_{\mu \nu} - \varepsilon_{\mu\nu} ))$ can be used. The last non trivial identity is \cref{eq:strain_omega}. There, the useful identity $\pdv{\det\vb M}{x} = \det \vb M \Tr (\vb M^{-1} \pdv{\vb M}{x})$ and the transformation law of the unit cell matrix $h_{\mu \nu}' = (\delta_{\mu \alpha} + \varepsilon_{\mu \alpha}) h_{\alpha \nu}$ is used.

The derivatives required form terms $\vb A$ and $\vb C$ are given in \cref{eq:strain_green,eq:strain_omega} respectively. Using the product rule, term $\vb B$ can be written as $\pdv{\abs{\Tilde{\rho}(\vb G)}^2}{\strain} = \pdv{\Tilde{\rho}(\vb G) \cdot \Tilde{\rho}^*(\vb G)}{\strain} = \Tilde \rho^*(\vb G) \pdv{\Tilde \rho (\vb G)}{\strain} + \Tilde \rho(\vb G) \pdv{\Tilde \rho ^* (\vb G)}{\strain} $ where $\pdv{\Tilde \rho (\vb G)}{\strain}$ is the Fourier transform of 
$\pdv{\rho(x)}{\strain}$. Because of \cref{eq:fourier_factor} $\pdv{e^{i\vb G \vb r}}{\strain} = 0$ and therefore strain derivatives commute with the Fourier transform.

\section{Charge density with periodic boundary conditions and its strain derivative}\label{asec:strain}

\begin{equation}
\rho(\vb r, \vb R_1, \dots, \vb R_{N_{\mathrm{at}}}, \vb h) = \sum_{i, j, k = -\infty}^\infty \sum_{l=1}^{N_{\mathrm{at}}} q_l \rho_l\left( \underbrace{ \norm{\vb r - \vb R_l - i \vb h_1 - j \vb h_2 - k \vb h_3  }^2}_{\vb{x}^2} \right)
\end{equation}
Only $\pdv{\rho_l(\vb x^2)}{\strain}$ needs to be calculated to get the strain derivative or the charge density.
\begin{equation}
    \pdv{\rho_l(x_\tau x_\tau)}{\strain} = \pdv{x_\tau x_\tau}{\strain} \rho_l'(\vb x^2) = 2 x_\tau \pdv{x_\tau}{\strain} \rho_l'(\vb x^2) = 2 x_\tau \delta_{\tau \mu} x_\nu \rho_l'(\vb x^2) = 2 x_\mu x_\nu \rho_l'(\vb x^2)
\end{equation}
Finally, 
\begin{equation} \label{aq:rho_pbc_general}
    \pdv{\rho(\vb r, \vb R_1, \dots, \vb R_{N_{\mathrm{at}}}, \vb h)}{\strain} =  \sum_{i, j, k = -\infty}^\infty \sum_{l=1}^{N_{\mathrm{at}}} 2 q_l ( \vb r - \vb R_l - i \vb h_1 - j \vb h_2 - k \vb h_3 )_\mu ( \vb r - \vb R_l - i \vb h_1 - j \vb h_2 - k \vb h_3 )_\nu \rho_l'(( \vb r - \vb R_l - i \vb h_1 - j \vb h_2 - k \vb h_3 )^2)
\end{equation}
or with Gaussian atomic charge densities $\rho_l(\norm{\vb x}^2) = \frac{1}{\sqrt{ 2 \pi \sigma_l^2 }^3} e^{- \frac{\norm{\vb x}^2}{2 \sigma_l ^2} }$
\begin{equation}
    \pdv{\rho(\vb r, \vb R_1, \dots, \vb R_{N_{\mathrm{at}}}, \vb h)}{\strain} = - \sum_{i, j, k = -\infty}^\infty \sum_{l=1}^{N_{\mathrm{at}}} \frac{q_l}{\sqrt{ 2 \pi \sigma_l^2 }^3} \frac{1}{\sigma_l^2} ( \vb r - \vb R_l - i \vb h_1 - j \vb h_2 - k \vb h_3 )_\mu ( \vb r - \vb R_l - i \vb h_1 - j \vb h_2 - k \vb h_3 )_\nu e^{- \frac{(\vb r - \vb R_l - i \vb h_1 - j \vb h_2 - k \vb h_3  )^2}{2 \sigma_l ^2} }.
\end{equation}

\section{Charge equilibration total derivatives}
\label{asec:totderiv}
Here, \cref{eq:force_int,eq:strain_part1} are simplified.
\begin{align}
        \sum_{i=1}^\nt \lambda_i \sum_{j=1}^\nt \pdv{A_{ij}}{\vb r_k} Q_j &= \sum_{i, j=1}^\nt \lambda_i q_j \pdv{}{\vb r_k} \int \int \frac{\rho_i(\vb r -\vb r_i) \rho_j(\vb r' - \vb r_j) }{\norm{\vb r - \vb r'}} d\vb r d\vb r' \nonumber \\
        &= \pdv{}{\vb r_k} \int \rho^{\bm{\lambda}}(\vb r) V^{\vb Q}(\vb r) d\vb r =
        \int \left[ V^{\vb Q}(\vb r) \pdv{\rho^{\bm \lambda}(\vb r)}{\vb r_k} + V^{\bm \lambda}(\vb r) \pdv{\rho^{\vb Q}(\vb r)}{\vb r_k} \right] d\vb r \label{asec:qeq_tot_deriv}
\end{align}

\begin{align} \nonumber
    &\sum_i \lambda_i \sum_j \pdv{A_{ij}}{\strain} Q_j  = \pdv{}{\vb \strain} \int \rho^{\bm{\lambda}}(\vb r) V^{\vb Q}(\vb r) d\vb r = 
    \pdv{}{\vb \strain} 2 \pi\Omega \sum_{\vb G\neq 0}  \rho^{\bm{\lambda} *}(\vb G) \Tilde V^{\vb Q}(\vb G) \\
    &= 2 \pi \Omega \sum_{\vb G\neq 0} \left[ \delta_{\mu\nu}  \rho^{\bm{\lambda} *}(\vb G) \Tilde V^{\vb Q}(\vb G) +  \pdv{ \tilde \rho^{\bm \lambda *}(\vb G)}{\strain} \tilde V^{\vb Q}(\vb G) + \tilde \rho^{\bm \lambda*}(\vb G) \pdv{ \tilde V^{\vb Q}(\vb G)}{\strain} \right]\nonumber\\
    &=  2 \pi \Omega \sum_{\vb G\neq 0} \left[ \delta_{\mu\nu}  \rho^{\bm{\lambda} *}(\vb G) \Tilde V^{\vb Q}(\vb G) +  \pdv{ \tilde \rho^{\bm \lambda *}(\vb G)}{\strain} \tilde V^{\vb Q}(\vb G) + \frac{\tilde \rho^{\bm \lambda*}(\vb G)}{\vb G^2} \left( \pdv{ \tilde \rho^{\vb Q}(\vb G)}{\strain} + \frac{2}{\vb G^2} G_\mu G_\nu  \tilde \rho^{\vb Q}(\vb G)\right) \right] \nonumber\\
    &=2 \pi \Omega \sum_{\vb G\neq 0} \frac{1}{\vb G^2} \left[ \delta_{\mu\nu}  \rho^{\bm{\lambda} *}(\vb G) \Tilde \rho^{\vb Q}(\vb G) +  \pdv{ \tilde \rho^{\bm \lambda *}(\vb G)}{\strain} \tilde \rho^{\vb Q}(\vb G) + \tilde \rho^{\bm \lambda*}(\vb G) \left( \pdv{ \tilde \rho^{\vb Q}(\vb G)}{\strain} + \frac{2}{\vb G^2} G_\mu G_\nu  \tilde \rho^{\vb Q}(\vb G)\right) \right] \label{aq:4g_stress_trick}
\end{align}

\end{widetext}

\bibliographystyle{apsrev4-2}
\bibliography{main}

\end{document}